# Suspended InAs nanowire Josephson junctions assembled via dielectrophoresis

Domenico Montemurro[1,2], Daniela Stornaiuolo[3], Davide Massarotti[3], Daniele Ercolani[1], Lucia Sorba[1], Fabio Beltram[1], Francesco Tafuri[2], Stefano Roddaro[1]

[1] NEST, Istituto Nanoscienze-CNR and Scuola Normale Superiore, Pisa I-56127, Italy
[2] Seconda Università degli Studi di Napoli, Dipartimento di Ingegneria dell'Informazione, Aversa (CE) I-81031,I Italy
[3] Università degli Studi di Napoli Federico II and CNR-SPIN, Napoli I *Via Cinthi*a 26, I-80126, Italy

**ABSTRACT**
We present a novel technique for the realization of suspended Josephson junctions based on InAs semiconductor nanowires. The devices are assembled using a technique of drop-casting guided by dielectrophoresis that allows to finely align the nanostructures on top of the electrodes. The proposed architecture removes the interaction between the nanowire and the substrate which is known to influence disorder and the orientation of the Rashba vector. The relevance of this approach in view of the implementation of Josephson junctions based on High-Temperature Superconductors is discussed.

**INTRODUCTION.** Hybrid devices based on low-dimensional materials coupled to superconducting (SC) electrodes emerged in the last years as an ideal platform for the investigation unconventional quantum effects. In these devices the macroscopic quantum correlations typical of SCs can be induced via proximity effect in a wide class of low-dimensional materials such as graphene[1,2] and semiconducting nanowires (NWs)[3-7] that offer a unique degree of freedom in the design and control of quantum states[8-11]. The most intriguing opportunity offered by these material combinations is the possibility to drive novel excitations and properties that none of the individual device components originally possessed. Recently, this possibility was at the base of the proposal to obtain Majorana-like excitations in semiconducting NWs[11-15], where pairing correlations can be combined with Rashba spin-orbit coupling[15-18]. Hybrid NWs-SC Josephson junctions are typically obtained by transferring NWs to a Si/SiO$_2$ substrate, on top of which device electrodes are nanofabricated by aligned e-beam lithography. Future advancements in the field and a better understanding of the exotic excitations induced in NW-SC hybrid structures will also depend on the possibility to realize devices having complex architectures, such as systems comprising several NWs connected in a controlled way[19]. Moreover, in these hybrid devices the NW-to-substrate interaction is known to play an important role in terms of induced disorder and by the breaking of inversion symmetry in the electrostatic environment in which the nanostructure is embedded. This is believed to lead to a pinning of the Rashba vector within the plane of the host Si substrate[20]. In this letter we demonstrate a rapid and effective method to realize NW-based suspended Josephson junctions where nanostructure positioning is obtained by dielectrophoresis (DEP) on previously patterned electrodes[21-23]. In our devices, single NWs are aligned on electrode pairs separated by nanogaps approx. $100nm$, corresponding to a dimension which is over one order of magnitude smaller than the NW length. This is an unusual and challenging DEP configuration and we discuss a novel geometry for achieving a good deposition in this regime. We also show that the properties of the hybrid devices realized using this method are comparable with those of conventional non-suspended devices. We believe that our DEP technique is suitable for the fabrication of a wide class of hybrid NW-based devices which can include, for instance, High critical Temperature Superconductors (HTS)[24]. The relevance of our protocol for the implementation of field-effect HTS hybrid devices will be discussed.

**DEVICE FABRICATION** The structure of the suspended devices we realized is shown in Fig.1.
Junctions are fabricated using highly-doped n-type InAs NWs grown by chemical beam epitaxy[25] and characterized by a charge density of n=1.8±0.8x $10^{19}$cm$^{-3}$, a diameter of 90±10nm and an average length of 2μm. The DEP electrodes are realized starting from a Ti/Au bilayer (10/100 nm) thermally evaporated on top of a SiO$_2$/Si substrate (280nm oxide) and patterned using a combination of UV and e-beam lithography. First, a set of bonding pads is defined by optical lithography, subsequently nanogaps are fabricated by e-beam lithography with a good reproducibility down to a gap size d≈140nm (Fig.1a). The design of the Ti/Au electrodes is a crucial aspect of our method, since a suitable geometry is required to efficiently attract the NWs by DEP[26]. Our optimal geometry is discussed in further details in the next Section. Following nanogap fabrication, NWs are placed across them using DEP-guided drop-casting (Fig.1b). In order to realize the SC contacts, a further e-beam lithographic step and thermal evaporation is used to define two 0.3 x 1 μm² Ti/Al (12/120nm) patches (Fig.1c) separated by a distance $L \propto d$, which will be regarded in the rest of the paper as the actual junction length. A good contact transparency between the metal and the NWs is ensured by a passivation step in an ammonium polysulfide ($NH_4S_x$) solution for 35 seconds at 44 °C prior to Ti/Al deposition. In parallel to the suspended junctions, conventional Al-NW weak links were fabricated also directly on top of the SiO₂/Si substrate in order to compare the properties of the devices in the two architectures. In this case the fabrication protocol was identical to the one used for the Ti/Al patches in the suspended devices.

## GUIDING NANOWIRE DEPOSITION BY DIELECTROPHORESIS

DEP is a well-known technique for the controlled deposition of nano- and micro-scale structures. It exploits the forces exerted on a dielectric particle by the non-uniform electric field **E** stemming from the induced dipole moment **p**. In particular, when the size of the nano-/micro-structure can be neglected, one can expect a force and torque given by

$$\mathbf{F} = (\mathbf{p}.\nabla)\mathbf{E} \qquad (1)$$
$$\mathbf{T} = \mathbf{p} \times \mathbf{E} \qquad (2)$$

The magnitude of the induced dipole **p** depends on the complex dielectric functions of the DEP liquid medium $\varepsilon^*_m(\omega)$ and of the particle $\varepsilon^*_p(\omega)$, where $\omega = 2\pi\nu$ (ν is the frequency of the AC electric field and $\varepsilon^* = \varepsilon - j\sigma\omega$). The resulting force can be readily calculated for the case of a prolate ellipsoid: it is proportional to the gradient of the electrostatic energy following the law $\mathbf{F} \propto \varepsilon^*_m(\omega)\mathrm{Re}[K_\alpha(\omega)]\nabla\mathbf{E}^2$ where we introduced the Clausius-Massotti factors[21]

$$K_\alpha[(\omega)] = \frac{\varepsilon^*_p - \varepsilon^*_m}{A_\alpha(\varepsilon^*_p - \varepsilon^*_m) + \varepsilon^*_m}, \qquad \alpha = L, S \qquad (3)$$

The form factors $A_\alpha$ depend on the eccentricity $e$ of the ellipsoid and attain two different values when the field is oriented along the long (L) or short (S) axes. Our NWs have a length which is ≈20 times the diameter; we can thus estimate $e = \sqrt{1 - (1/20)^2}$,
leading[21] to $A_L \approx 0.007$ and $A_S \approx 0.497$. The positive or negative sign of $\mathrm{Re}[K_\alpha]$ determines whether the particle is attracted (positive DEP) or repelled (negative DEP) from regions with large **E**. Since our DEP protocol was performed using isopropyl alcohol ($\varepsilon_m \approx 18$ and negligible conductivity $\sigma_m \approx 6\mu S/m$) as the medium and highly-doped InAs NWs as the ``particle'' ($\varepsilon_p \approx 10$ and sizable conductivity $\sigma_p \approx 10^5 S/m$), deposition is expected to work already at rather low frequencies[27]. Provided the correct DEP sign is achieved, the force in equation (1) will tend to attract the nanostructure to regions with a large electric field, the torque in equation (2) will tend to align the NWs to the field lines thanks to their easier polarizability along the axis. Usually, DEP is performed using electrodes separated by a gap of size comparable to that of the target nanostructures. In our case, on the other hand, we need to position the NWs on top of electrodes separated by a gap with a size d approx. 100-200nm, thus 10-20 times smaller than the NW length; this is an unusual configuration for DEP. One issues with such a small-gap geometry is that the field **E** in the gap can become very strong even for relatively moderate excitation voltages, with a significant risk of destructive electrostatic discharges. This poses strict limitations to the highest voltage that can be safely applied to the electrodes and, in turn, reduces the size of the region where DEP forces are able to overcome the effect of Brownian motion and random liquid convection and thus to actually capture NWs. In addition, for small structures, trapping forces converging to the gap occur in a region which can easily be even smaller than the

NW size and rapidly decay in the surrounding volume; therefore they are not very effective in trapping the nanostructure. These issues motivated us to develop a new DEP electrode geometry. In order to obtain a four-wire contact geometry, we designed two facing electrodes with a ``U'' shape and large surface coverage so that a sizable **E** field could be induced over a significant volume and lead to a larger capture cross-section. Prior to the fabrication, we performed numerical simulations of the induced DEP field for structures of various dimensions. In Fig.2 we compare predictions[28] for a standard case of two 1μm-wide fingers separated by a thin gap of 200nm (panel a) and for a w=10μm gap geometry (panel b), which reproduce the experimental configuration of Fig.1e. In both cases we plot the modulus of the gradient of the electrostatic energy density for a biasing voltage of ±1V at the two electrodes and using $\varepsilon_m$=18.23 for the isopropyl alcohol. A 2μm-long NW was added as a length reference. The comparison indicates that the second geometry is more effective in capturing the DEP is a well-known technique for the controlled deposition of nano- and micro-scale structures. It exploits the forces exerted on a dielectric particle by the non-uniform electric field **E** stemming from the induced dipole moment **p**. In particular, when the size of the nano-/micro-structure can be neglected, one can expect a force and torque given by

$$\mathbf{F} = (\mathbf{p}.\nabla)\mathbf{E} \qquad (1)$$
$$\mathbf{T} = \mathbf{p} \times \mathbf{E} \qquad (2)$$

The magnitude of the induced dipole **p** depends on the complex dielectric functions of the DEP liquid medium $\varepsilon^*_m(\omega)$ and of the particle $\varepsilon^*_p(\omega)$, where $\omega = 2\pi\nu$ ($\nu$ is the frequency of the AC electric field and $\varepsilon^* = \varepsilon - j\sigma\omega$). The resulting force can be readily calculated for the case of a prolate ellipsoid: it is proportional to the gradient of the electrostatic energy following the law $\mathbf{F} \propto \varepsilon^*_m(\omega)\text{Re}[K_\alpha(\omega)]\nabla\mathbf{E}^2$ where we introduced the Clausius-Massotti factors[21]

$$K_\alpha[(\omega)] = \frac{\varepsilon^*_p - \varepsilon^*_m}{A_\alpha(\varepsilon^*_p - \varepsilon^*_m) + \varepsilon^*_m}, \qquad \alpha=L,S \qquad (3)$$

The form factors $A_\alpha$ depend on the eccentricity $e$ of the ellipsoid and attain two different values when the field is oriented along the long (L) or short (S) axes. Our NWs have a length which is ≈20 times the diameter; we can thus estimate $e = \sqrt{1 - (1/20)^2}$,
leading[21] to $A_L \approx 0.007$ and $A_S \approx 0.497$. The positive or negative sign of $\text{Re}[K_\alpha]$ determines whether the particle is attracted (positive DEP) or repelled (negative DEP) from regions with large **E**. Since our DEP protocol was performed using isopropyl alcohol ($\varepsilon_m \approx 18$ and negligible conductivity $\sigma_m \approx 6\mu S/m$) as the medium and highly-doped InAs NWs as the ``particle'' ($\varepsilon_p \approx 10$ and sizable conductivity $\sigma_p \approx 10^5 S/m$), deposition is expected to work already at rather low frequencies[27]. Provided the correct DEP sign is achieved, the force in equation (1) will tend to attract the nanostructure to regions with a large electric field, the torque in equation (2) will tend to align the NWs to the field lines thanks to their easier polarizability along the axis. Usually, DEP is performed using electrodes separated by a gap of size comparable to that of the target nanostructures. In our case, on the other hand, we need to position the NWs on top of electrodes separated by a gap with a size d approx. 100-200nm, thus 10-20 times smaller than the NW length; this is an unusual configuration for DEP. One issues with such a small-gap geometry is that the field **E** in the gap can become very strong even for relatively moderate excitation voltages, with a significant risk of destructive electrostatic discharges. This poses strict limitations to the highest voltage that can be safely applied to the electrodes and, in turn, reduces the size of the region where DEP forces are able to overcome the effect of Brownian motion and random liquid convection and thus to actually capture NWs. In addition, for small structures, trapping forces converging to the gap occur in a region which can easily be even smaller than the NW size and rapidly decay in the surrounding volume; therefore they are not very effective in trapping the nanostructure. These issues motivated us to develop a new DEP electrode geometry. In order to obtain a four-wire contact geometry, we designed two facing electrodes with a ``U'' shape and large surface coverage so that a sizable **E** field could be induced over a significant volume and lead to a larger capture cross-section.

Prior to the fabrication, we performed numerical simulations of the induced DEP field for structures of various dimensions. In Fig.2 we compare predictions[28] for a standard case of two 1µm-wide fingers separated by a thin gap of 200nm (panel a) and for a w=10µm gap geometry (panel b), which reproduce the experimental configuration of Fig.1e. In both cases we plot the modulus of the gradient of the electrostatic energy density for a biasing voltage of ±1V at the two electrodes and using $\varepsilon_m$=18.23 for the isopropyl alcohol. A 2µm-long NW was added as a length reference. The comparison indicates that the second geometry is more effective in capturing the NWs and orienting them along the planned direction, for a given electrode bias since DEP forces extend further away from the gap with respect to the case of panel (a). This is even more evident looking at the white lines on the right-hand side plots, which indicate the direction of the DEP action: in our DEP geometry attractive forces converge to the gap over a rather large volume; differently, as anticipated, a much poorer capture efficiency can be expected in the more standard geometry of panel (a). Experimentally, our DEP process was achieved as follows. InAs NWs were dispersed in isopropanol by sonication of the growth substrate. The deposition was then achieved by putting a drop (2µl) of the solution on the substrate and, at the same time, by applying an AC voltage to the Ti/Au electrodes, as indicated in the overlay in Fig.1d. Since we use a very small inter-electrode gap, the substrate is largely screened. As a consequence, it is not expected to have an important effect and can be left floating[21]. Optimal results were obtained by applying an AC voltage in the range 0.75 - 1.20V and a frequency of 1-10kHz, while frequencies $\nu \leq$ 10Hz were found to be less effective[21-23]. The AC field was applied until the isopropanol completely dried up. These values gave a good deposition yield, without accumulation of NW clusters. A typical device obtained using this process is shown in Fig.1e.

## ELETRICAL TRANSPORT MEASUREMENTS

Devices fabricated following the DEP protocol described in the previous section yielded Josephson junctions with properties which were found to be comparable to those typically observed in the more standard non-suspended architecture. The transport properties of the devices were measured from room temperature down to 300mK using a Heliox refrigerator. The measurement system was magnetically shielded, using superconducting screens, and the lines were filtered using RC filters and two stages of copper-powder filters. Figure 3 shows typical Resistance vs. Temperature (R(T)) curves of suspended devices fabricated starting from nano-trenches of different sizes d (see Fig.1e) and having a channel length L≈d. Since the room-temperature normal resistance $R_N$ of the devices ranges from 180 to 400Ω, the plot shows the resistances normalized to their values at T=1.4K. The curves were recorded using a lock-in technique with an excitation current of 5nA. From room temperature down to T=1.1K the R(T) curves show a similar insulating trend (data not shown) independently from the extension L of the normal region. For all devices, a drop in the resistance is observed at T=1.1K, as a consequence of the superconducting transition of the Ti/Al pads. Below this temperature, we observed a different behavior depending on the L parameter. For devices with L≤150nm, a complete superconducting transition was typically recorded. These results clearly indicate the influence of the channel length L on the proximity effect across the contact barriers and into the NWs.

The current vs. voltage (I-V) characteristics of the superconductive devices were carefully analyzed as a function of temperature: selected curves are shown in Fig.4a for device D7B. The behavior of this devices is typical of that of an overdamped Josephson junction showing phase-diffusion effects that become particularly evident for T>400mK[29,30]. The critical current, extracted using a V= 1µV threshold, is $I_c$=62nA at T=300mK; this value corresponds to a critical-current density of about 1.0kA/cm$^2$, in agreement with the values typically found for this type of InAs NWs[7]. These superconducting properties are comparable to those of conventional devices fabricated directly on the Si/SiO$_2$ substrate. The I-V characteristic of one of these conventional devices (D7T) is reported in Fig.4b and exhibits an $I_c$ value at 300mK similar to the one obtained in suspended device D7B. This comparison confirms that the discussed DEP method provides a valid route for the realization of high quality hybrid devices on narrow-gap electrodes. Starting from the normal transport properties of the InAs NWs, we estimate a diffusion coefficient D=0.02m$^2$/s, a Thouless Energy $E_{Th} = \hbar D/L^2$ and a normal coherence length $\xi_N = \sqrt{\hbar D/(2\pi k_B T)}$=300nm; these parameters indicate that the devices operate in the diffusive short-junction limit. Further insight in the transport mechanisms of the suspended devices can be obtained from the temperature dependence of the Josephson critical current

$I_c(T)$, which we show in Fig.4c. We compare the experimental $I_c$ vs. T behavior with a model which takes into account the barrier transparency. The upper concavity of the $I_c$ vs T curve of Fig.4c is indeed an indication of a sizable superconducting (S)-normal (N) interface resistance, as shown in Ref.31}. In this framework, the energy scale ruling the $I_c(T)$ dependence is an effective Thouless Energy $E_{Th}*$ which takes into account the ratio between the resistances of the S/N interfaces and the one of the normal conductor. Our experimental data can be well approximated by an exponential function[32]:

$$I_C \propto \exp(-T/T^*) \qquad (4)$$

where $T^*$ is linked to the reduced Thouless energy by $E_{Th}*=2\pi k_B T^*/24$. The fit is shown as a solid red line in Fig.4c and was obtained with a value $E_{th}*=3.1\mu eV$, confirming the effect of the interface resistance in hampering the superconducting proximization of the NW. Finally, in Fig.5 we show the temperature evolution of the differential resistance dV/dI curves as a function of the voltage in the sub-gap region (considering $\Delta=120\mu eV$ for Al), as visible in Fig.5. Symmetric structures are observed at V≈10 and 25μeV. The voltage position of these peaks is not influenced by temperature and suggests they cannot be linked to multiple Andreeev reflections. However, the amplitude of the peaks decreases as the temperature is increased and is almost completely suppressed at T=720mK: this evolution indicates that these structures do have some connection with superconductivity. Several works in the literature reported the appearance of differential resistance peaks having similar temperature behavior and related them to the coupling of Josephson oscillations to the acoustic degrees of freedom of the suspended NWs[33] However, since we observed similar peaks also in non-suspended devices (inset of Fig.5), we believe this interpretation is not applicable to our case. S. Abay and coworkers in Ref.34 suggested an alternative mechanism based on the presence of inhomogeneities in the NW. In this scenario, the observed peaks would be related to the sequential appearance of normal domains in the NW as the current is increased. The presence of an inhomogeneous resistive state in the NW region in proximity with Al is supported in our case by the peculiar shape of the R(T) curves. As visible from the data of Fig.3, the R(T) curve of devices with L in the range 150 - 160nm exhibit a large plateau which extends from the superconductive onset of the Al patches at T=1.1K, down to T=500mK.

**CONCLUSIONS** In conclusion, we have demonstrated a novel DEP technique for the fabrication of suspended Josephson junctions. Since the inter-electrode gap in our device architecture is much smaller than the NW dimension, a novel DEP geometry was simulated and experimentally tested. The possibility to decouple the NW from the $SiO_2$ substrate is a desirable feature since it is expected to lead to a reduced scattering and to avoid the symmetry breaking effects stemming from NW-substrate interactions. The comparison between suspended and non-suspended devices indicates that the fabrication protocol is robust. We note that in our protocol the superconductive electrodes can be fabricated before NW deposition. This is a crucial difference which can enable to use of superconductors whose growth/deposition conditions are not compatible with the chemical and structural stability of the NWs. In particular it could be combined with HTS layers which require rather extreme growth conditions. The fabrication of HTS-NW Josephson junctions could be relevant to the observation of Majorana excitations in hybrid systems[24], allow higher operation temperatures and a more robust behavior in external magnetic fields.

## Acknowledgements

This work was supported by the Italian MIUR FIRB project *HybridNanoDev RBFR1236VV001* and by *POR Campania FSE 2007*-2013 "MAteriali e STrutture Intelligenti" MASTRI.

# Figures

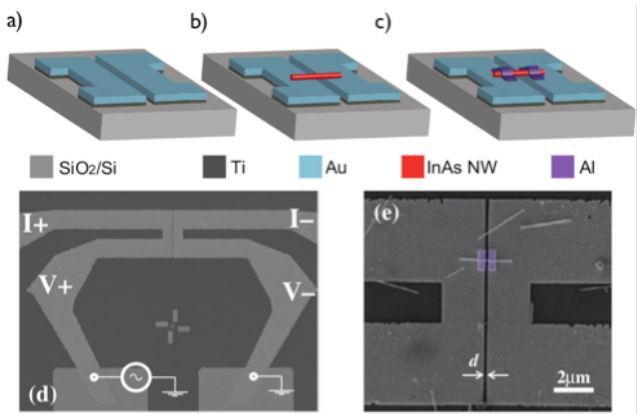

**Figure 1** A sketch of the fabrication procedure is visible in the top panels: Ti/Au electrodes (blue) with a nanotrench are fabricated by a combination of e-beam and optical lithography (a), InAs nanowires (red) are deposited across the inter-electrode gap by a dielectrophoresis technique (b) and contacted with two 0.3 x 1 μm² superconductive Ti/Al patches (violet) (c). Panel (d) shows the layout of the Ti/Au contact geometry, the overlay indicates the AC bias system used to achieve a deposition guided by dielectrophoresis. Panel (e) shows an example of a complete device, with the Ti/Al patches in violet.

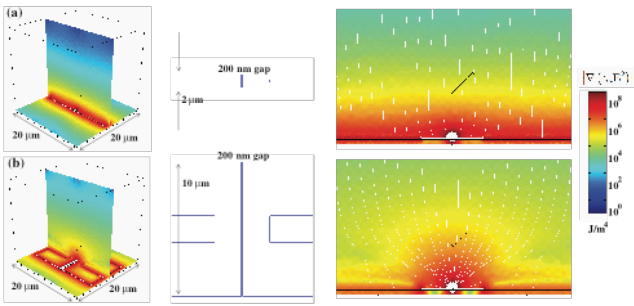

**Figure 2** Numerical simulations for the gradient of the electrostatic energy density $|\nabla \varepsilon \mathbf{E}^2/2|$ for two different electrode configurations. Panel (a) refers to double finger architecture with a lateral size of 1 μm and a gap of 200 nm. Panel (b) refers to the geometry shown in Fig.1. In both cases a ±1V bias was applied to the two electrodes and using $\varepsilon_m$=18.23 for the isopropyl alcohol. On the right had side white lines indicate the direction of the gradient and highlight the more efficient trapping of the geometry in panel (b). It is important to note that the field line density in the right panels does not have any direct correspondence with $|\mathbf{E}|$ since the field configuration is three-dimensional and has a strong dependence on the out-of-plane direction, particularly in thin-finger the case (a). A 2μm long cylinder was added as a reference at about 3μm from the electrodes.

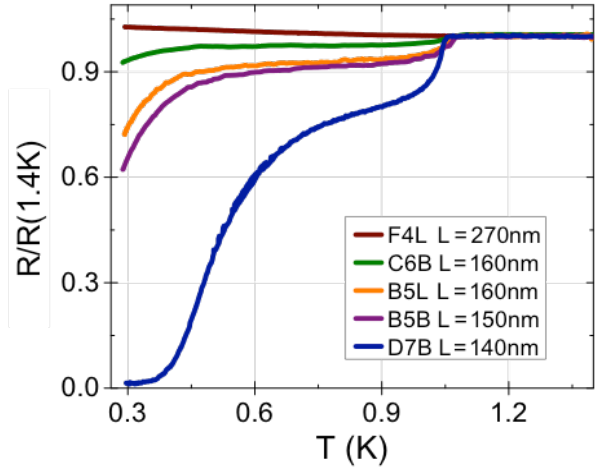

**Figure 3** Normalized R(T)/R(1.4K) curves of suspended devices characterized by a different channel length L. A different behavior is observed as a function of L: devices with L<200nm display a metallic behavior instead if L<150nm a full transition to the superconductive regime is obtained.

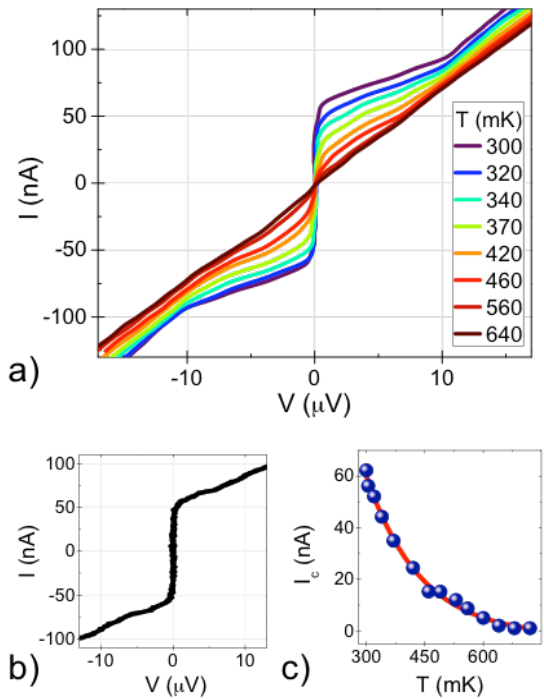

**Figure 4** I-V characteristics of suspended device D7B as a function of temperature (a) and of conventional non-suspended device D7T at T=300mK (b). The two devices have the channel length L=140nm. The Ic vs T of device D7B (blue dots) is visible in panel (c). The full red line is the fit obtained using equation 4.

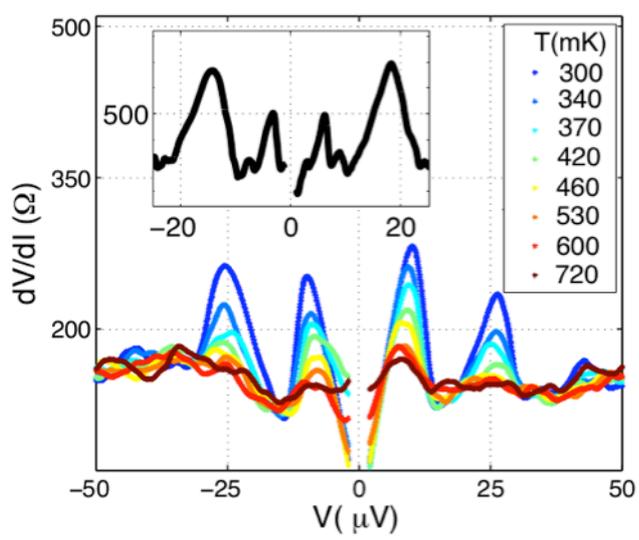

**Figure 5** Differential resistance dV/dI as a function of V of suspended device D7B for selected temperatures. The low voltage part of the data (where superconductivity is present) was cut for clarity. Only the sub-gap region is shown in order to highlight the presence of peaks with temperature independent voltage position. Similar peaks are visible also in the dV/dI curve of non-suspended device D7T, which is shown in the inset.